\documentclass[12pt]{article} 
\usepackage{amsmath}
\usepackage{amssymb}
\begin{document}  
\title{\bf Ambiguities on the Hamiltonian formulation of the free falling  particle with quadratic dissipation}
\author{ G. L\'opez$^{1}$, P. L\'opez$^{1}$, X. E. L\'opez$^{2}$, and R. I. Castro$^{1}$\\ \\
\small $^{1}$ Departamento de F\'{\i}sica de la Universidad de Guadalajara\\
\small Apartado Postal 4-137\\
\small 44410 Guadalajara, Jalisco, M\'exico\\ \ \\
\small $^{2}$ Facultad de Ciencias de la UNAM\\
\small Apartado Postal 70-348, Coyoac\'an  04511 M\'exico D.F.}  

\maketitle

\begin{abstract}
For a free falling particle moving in a media which has quadratic velocity force effect on the particle,
two equivalent constants of motion, with units of energy, two Lagrangians, and two Hamiltonians are deduced.
These quantities describe the dynamics of the same classical system. However, their quantization and the associated 
statistical mechanics (for an ensemble of particles) describe two completely different quantum and statistical systems. 
This is shown at first order in the dissipative parameter.
\end{abstract}


\section{\label{I} Introduction}

It is well known that the Lagrangian (therefore the Hamiltonian) formulation for some systems of more than one dimension 
may not exists (Douglas 1941). Fortunately for our study of the nature up to now, most of our physical systems have avoided
this problem, and the whole quantum and statistical mechanics of non-dissipative systems can be given in terms of a
Lagrangian or Hamiltonian formulation. Now, for dissipative systems there have been two main approaches. The first one
consists of keeping the same Hamiltonian formalism for the whole system where the interacting background is included, as a
result, one brings about a master equation with the dissipation and diffusion parameters appearing as part of the solution
(Caldeira and Legget 1983, Unruh and Zurek 1989, and Hu et al 1992). This approach has its own merit, but it will not be
followed in this paper. We will follow the second approach  which consists in to obtain a phenomenological velocity depending
Hamiltonian, representing a classical dissipative system, and to proceed to make the usual quantization (or statistical
mechanics) with this Hamiltonian.

Within this last approach, one can, additionally, study the mathematical consistence of the the Hamiltonian formalism in 
quantum and statical mechanics. It is also known that even for one-dimensional systems, where the existence of their
Lagrangian is guaranteed (Darboux 1894), the Lagrangian and Hamiltonian formulations are not free from problems (Havas 1973,
Okubo 1980, Dodonov et al 1981, Marmon et al 1985, Glauber et al 1984, $^a$L\'opez 1998, and $^b$L\'opez 1999). One of the
main problems is the implication on the quantization of the associated classical system  when different Hamiltonians describe
the same classical system ($^c$L\'opez 2002). This ambiguity has already been studied for the harmonic oscillator with
dissipation and some general system ($^d$L\'opez 1996). In this paper, we want to show explicitly this ambiguity by studying
the free falling particle within a medium which has the effect on the particle of producing the dissipation. This dissipation
depends quadratically on  the velocity of the particle. Firstly, two constants of motion are deduced for this system.
Secondly, with these constants of motion two Lagrangian and two Hamiltonian are obtained using a known procedure ($^d$L\'opez
1996 and $^e$L\'opez and Hern\'andez 1989). Finally, using the Hamiltonian expression at first order in the dissipation
parameter, the resulting eigenvalues of their associated quantum Hamiltonian and their associated statistical mechanics 
properties (for an ensemble of particles) are shown. 
\vfil\eject\noindent

\section{\label{M} Constants of Motion}

The motion of the particle of mass $m$ falling under a constant gravitational force, $-mg$, where $g$ is the constant 
acceleration due to gravity, which is within a dissipative medium which has the effect on the particle of producing a force
proportional to the square of the velocity of the particle, $\alpha \dot x^2$ for $\dot x<0$, can be described by the
following autonomous dynamical system
\begin{equation}
\frac{dx} {dt}=v\ ,\hskip2cm \frac{dv}{ dt}=-g+\frac{\alpha} {m}v^2\ ,
\label{I:defS}\end{equation}
where the variable $x$ represents the vertical position of the particle, and $v$ represents its velocity. A constant of 
motion for this system is a function $K=K(x,v)$ such that $dK/dt=0$, i.e. it satisfies the following equation ($^f$L\'opez
1999)
\begin{equation}
v\frac{\partial K}{\partial x}+\left(-g+\frac{\alpha}{ m}v^2\right)\frac{\partial K}{\partial v}=0\ .
\label{I:defK}\end{equation}
The general solution of this equation is given by (John 1974)
\begin{equation}
K_{\alpha}(x,v)=G(C(x,v))\ ,
\label{I:Kalpha}\end{equation}
where $G$ is an arbitrary function of the characteristic curve $C(x,v)$. This characteristic curve can be given in two 
different ways as
\begin{equation}
C_1=-\frac{mg}{ 2\alpha}\ln\left(1-\frac{\alpha} {mg}v^2\right)+gx\ ,
\label{I:C1}\end{equation}
or
\begin{equation}
C_2=\left(1-\frac{\alpha}{ mg}v^2\right)e^{-2\alpha x/m}\ .
\label{I:C2}\end{equation}
Considering that one must obtain the usual constant of motion (Energy) expression for $\alpha$ equal to zero, the 
functionality of $G$ in Eq. (3) is determined for each above characteristic ($G(C_1)=mC_1$ and $G(C_2)=-(m²
g/2\alpha)C_2-m^2g/2\alpha$), and the following constants of motion are gotten

\begin{equation}
K_{\alpha}^{(1)}(x,v)=-\frac{m^2g}{ 2\alpha}\ln\left(1-\frac{\alpha}{ mg}v^2\right)+mgx\ ,
\label{I:K1}\end{equation}
and

\begin{equation}
K_{\alpha}^{(2)}(x,v)=\frac{m^2}{ 2\alpha}\left(-g+\frac{\alpha}{ m}v^2\right)e^{-2\alpha x/m}+ \frac{m^2g}{ 2\alpha}\ . 
\label{I:K2}\end{equation}
Note that the following limit is gotten

\begin{equation}
\lim_{\alpha \to 0}K_{\alpha}^{(i)}=\frac{1}{ 2}mv^2+mgx\ \ \ i=1,2\ .
\label{I:Limits}\end{equation}
\section{\label{S} Lagrangians and Hamiltonians}
Using the know expression (Kobussen 1979, Leuber 1987, Yan 1981, and $^d$L\'opez 1996),

\begin{equation}
L(x,v)=v\int\frac{K(x,v)}{ v^2}dv\ ,
\label{II:L}\end{equation}
to get the Lagrangian, the Lagrangian associated to Eq.~(\ref{I:K1}) and Eq.~(\ref{I:K2}) are

\begin{equation}
L_{\alpha}^{(1)}(x,v)=
m\sqrt\frac{mg}{\alpha}~v~ arc\tanh\left(\sqrt\frac{\alpha}{ mg}~v\right)+
\frac{m^2g}{ 2\alpha}\ln\left(1-\frac{\alpha}{ mg}v^2\right)-mgx
\label{II:L1}\end{equation}
and

\begin{equation}
L_{\alpha}^{(2)}(x,v)=
\frac{m^2}{ 2\alpha}\left(g+\frac{\alpha}{ m}v^2\right)e^{-2\alpha x/m}-\frac{m^2g}{ 2\alpha}\ .
\label{II:L2}\end{equation}
Their generalized linear momenta ($p=\partial L/\partial v$) are

\begin{equation}
p_{\alpha}^{(1)}=m\sqrt\frac{mg}{\alpha}~arc\tanh\left(\sqrt\frac{\alpha}{ mg}~v\right)
\label{II:p1}\end{equation}
and

\begin{equation}
p_{\alpha}^{(2)}=mv e^{-2\alpha x/m}\ .
\label{II:p2}\end{equation}
Thus, their associated Hamiltonians, $H(x,p)=K(x,v(x,p))$, are given by

\begin{equation}
H_{\alpha}^{(1)}(x,p)=-\frac{m^2g}{ 2\alpha}\ln\left[1-\tanh^2\left(\sqrt\frac{\alpha}{ mg}~
\frac{p}{ m}\right)\right]+mgx
\label{II:H1}\end{equation}
and

\begin{equation}
 H_{\alpha}^{(2)}(x,p)=\frac{m^2}{ 2\alpha}\left(-g+\frac{\alpha p^2}{ m^3}e^{4\alpha x/m}\right)e^{-2\alpha x/m}+
\frac{m^2g}{ 2\alpha}\ ,
\label{II:H2}\end{equation}
where one has made the substitution of  $p_{\alpha}^{(1)}$ and $p_{\alpha}^{(2)}$ by
just $p$. One must note that the following limits are gotten

\begin{equation}
\lim_{\alpha\to 0}L_{\alpha}^{(i)}(x,v)=\frac{1}{ 2}mv^2-mgx\ ,
\label{II:LimL}\end{equation}

\begin{equation}
\lim_{\alpha\to 0}p_{\alpha}^{(i)}=mv\ ,
\label{II:Limp}\end{equation}
and

\begin{equation}
\lim_{\alpha\to 0}H_{\alpha}^{(i)}(x,p)=\frac{p^2}{ 2m}+mgx\hskip2cm i=1,2\ .
\label{II:LimH}\end{equation}
At  first order in the dissipation parameter $\alpha$, the Hamiltonians are given by

\begin{equation}
 H^{(1)}(x,p)=\frac{p^2}{ 2m}+mgx-\frac{\alpha}{ 12m^4g}p^4
\label{II:H1F}\end{equation}
and

\begin{equation}
H^{(2)}(x,p)=\frac{p^2}{ 2m}+mgx+\alpha\left(\frac{xp^2}{ m^2}-gx^2\right)\ .
\label{II:H2F}\end{equation}

\section{\label{F} Quantization at first order in perturbation theory}

Our Hamiltonians Eq.~(\ref{II:H1F}) and Eq.~(\ref{II:H2F}) can be written as

\begin{equation}
H^{(i)}(x,p)=H_0(x,p)+W^{(i)}(x,p)\hskip2cm\hbox{for}\hskip1cm i=1,2\ ,
\label{II:HQ}\end{equation}
where $H_0$ represents the non dissipative part of the Hamiltonian,
\begin{equation}
H_0(x,p)=\frac{p^2}{ 2m}+mgx\ ,
\label{II:H0Q}\end{equation}
and $W^{(i)}$ represents the contribution of the dissipation at first order in $\alpha$,

\begin{equation}
W^{(1)}(x,p)=-\frac{\alpha p^4}{ 12 gm^4}
\label{II:W1}\end{equation}
and

\begin{equation}
W^{(2)}(x,p)=\alpha\left(\frac{xp^2}{ m^2}-gx^2\right)\ .
\label{II:W2}\end{equation}
The Schr\"odinger equation, $i\hbar\partial\Psi(x,t)/\partial t=\hat{H}^{(i)}(\hat{x},\hat{p})\Psi(x,t)$ represents an 
stationary problem. Therefore, in order to get the quantization of the system, one just need to solve the following
eigenvalue problem

\begin{equation}
\hat{H}^{(i)}\Phi_n(x)=E_n^{(i)}\Phi_n(x)\ ,
\label{II:Eigen}\end{equation}
where $\hat{H}^{(i)}$ is the associated Hermitian operator of Eq.~(\ref{II:HQ}). Of course, one must not allow the 
particle to go  beyond down the surface level. Thus, Eq.~(\ref{II:H0Q}) is representing the Hamiltonian of the quantum bouncer
($x>0$) (Gean-Banacloche 1999), where the eigenvalue problem

\begin{equation}
\hat{H}_0(\hat{x},\hat{p})\psi_n^{(0)}(x)=E_n^{(0)}\psi_n^{(0)}(x)
\label{II:BH0}\end{equation}
has the eigenvectors and eigenvalues solution given by

\begin{equation}
\psi_n^{(0)}(x)=\frac{Ai(z-z_n)}{ |Ai'(-z_n)|}
\label{II:Bphi}\end{equation}
and

\begin{equation}
E_n^{(0)}=mgl_gz_n\ .
\label{II:BEn}\end{equation}
The functions $Ai$ and $Ai'$ are the Airy function and its differentiation respect to $z$. The variable $z$ is defined  as
$z=x/l_g$, where $l_g$ is given by $l_g=(\hbar^2/2m^2g)^{1/3}$, and $z_n$ is the nth-zero of the Airy function
($Ai(-z_n)=0$). In fact, the bouncing problem has already been studied for linear and quadratic dissipation ($^g$L\'opez
2004). For the later, the correction given to the eigenvalue problem using Eq.~(\ref{II:W2}) at first order in perturbation
theory is

\begin{equation}
\langle n|\hat{W}^{(2)}|n\rangle=\alpha\frac{4gl_g^2z_n^2}{ 15}\ .
\label{II:nW2n}\end{equation}
Now, using the relation $\langle n|d^4/dz^4|n\rangle=z_n^2/5$, the correction at first order in perturbation due to 
Eq.~(\ref{II:W1}) is given by

\begin{equation}
\langle n|\hat{W}^{(1)}| n\rangle=-\alpha\frac{\hbar^4z_n^2}{ 60 g m^4l_g^4}\ .
\label{II:nW1n}\end{equation}
Therefore, for the same classical dynamical system we have two different associated quantum systems which have completely 
different quantum dynamics, which is shown through the eigenvalues 

\begin{equation}
E_n^{(1)}=E_n^{(0)}+ \alpha\frac{4gl_g^2z_n^2}{ 15}
\label{II:E1}\end{equation}
and

\begin{equation}
E_n^{(2)}=E_n^{(0)}-\alpha\frac{\hbar^4z_n^2}{ 60 g m^4l_g^4}\ .
\label{II:E2}\end{equation}

\section{\label{G} Classical Statistical model for dissipation }

Consider a system of $N=N_1+N_2$ particles, where $N_1$ particles are small of mass $m_1$, and $N_2$ particles are are  big
of mass $m_2$ ($m_2\gg m_1$). The small particles move under the action of an external force with components $(0,0,-mg)$ and
suffer collisions with the walls of the container which consists in a narrow-square shape pipe of cross sectional area $L^2$.
In addition, each small particle can have occasional (stochastic) collision with the big particles, when they are added,
establishes the dissipative medium  where the big particles will move. The big particles move in this dissipative medium, and
it is assumed that, since this type of collision does not occur frequently, its average effect may have neglected
contribution on the dynamical macroscopic variables of the system. Newton's equations of motion for this system can be
written as

\begin{equation}
m_1\ddot{q}_{1ij}=0\hskip1cm j=1,\dots,N_1;\ i=x,y
\label{III:e1}\end{equation}

\begin{equation}
m_1\ddot{q}_{1zj}=-m_1g\hskip1cm j=1,\dots,N_1
\label{III:e2}\end{equation}

\begin{equation}
m_2\ddot{q}_{2ik}=\alpha\left(\dot{q}_{2ik}\right)^2\hskip1cm k=1,\dots,N_2;\ i=x,y
\label{III:e3}\end{equation}

\begin{equation}
m_2\ddot{q}_{2zk}=\alpha\left(\dot{q}_{2zk}\right)^2-m_2g\hskip1cm k=1,\dots,N_2\ ,
\label{III:e4}\end{equation}
where $q_{aij}$, $\dot{q}_{aij}$ and $\ddot{q}_{aij}$ are the generalized coordinates, velocities and accelerations of  the
light-small ($a=1$) and heavy-gross ($a=2$) particles, and the parameter $\alpha$ characterizes  the dissipative medium. The
Hamiltonian associated the the motion of 1-particle, Eq.~(\ref{III:e1}) and Eq.~(\ref{III:e2}), is given by ($^h$L\'opez et
al 1997)

\begin{equation}
H_{1;x,y,z}=\sum_{j=1}^{N_1}\left[\sum_{i=1}^3\frac{p_{1ij}^2}{ 2m_1}+m_1gq_{1zj}\right]\ .
\label{III:H1xyx}\end{equation}
The Hamiltonian associated to Eq.~(\ref{III:e3}) is written as

\begin{equation}
H_{2;x,y}=\sum_{k=1}^{N_2}\sum_{i=1}^2\frac{p_{2ik}^2}{ 2m_2}\exp\left(\frac{2\alpha q_{2ik}}{ m_2}\right) \ ,
\label{III:H2xy}\end{equation}
and, as we have seen in section 3, there are at least two Hamiltonians associated to Eq.~(\ref{III:e4}) which are given by

\begin{equation}
H_{2;z}^{(1)}=\sum_{k=1}^{N_2}\left\{-\frac{gm_2^2}{ 2\alpha}\ln\left[1-\tanh^2\left(\sqrt\frac{\alpha}{ m_2g}~
\frac{p_{2zk}}{ m_2}\right)\right]+m_2gq_{2zk}\right\}
\label{III:H12z}\end{equation} 
and
\begin{equation}
H_{2;z}^{(2)}=\sum_{k=1}^{N_2}\left \{ \frac{p_{2zk}^2}{ 2m_2}\exp\left(\frac{2\alpha q_{2zk}}{ m_2}\right)
-\frac{m_2^2g}{ 2\alpha}\left[\exp\left(-\frac{2\alpha q_{2zk}}{ m_2}\right)-1\right]\right\}\ .
\label{III:H22z}\end{equation}
Therefore, one has two different Hamiltonians to describe the same system, $H^{(1)}=H_{1;x,y,z}+H_{2;x,y}+H_{2;z}^{(2)}$ and 
$H^{(2)}= H_{1;x,y,z}+H_{2;x,y}+H_{2;z}^{(1)}$, which are written as
\begin{eqnarray*}
& & H^{(1)}=\sum_{j=1}^{N_1}\left[\sum_{i=1}^3\frac{p_{1ij}^2}{ 2m_1}+m_1gq_{1zj}\right]
 +\sum_{k=1}^{N_2}\sum_{i=1}^3\frac{p_{2ik}^2}{ 2m_2}\exp\left(\frac{2\alpha q_{2ik}}{ m_2}\right)\ \\ \
& & +\sum_{k=1}^{N_2}\left\{ \frac{p_{2zk}^2}{ 2m_2}\exp\left(\frac{2\alpha q_{2zk}}{ m_2}\right)
 -\frac{m_2^2g}{ 2\alpha}\left[\exp\left(-\frac{2\alpha q_{2zk}}{ m_2}\right)-1\right]\right\}
\end{eqnarray*}
\begin{equation}
\label{III:H1T}\end{equation} 
and
\vfil\eject
\begin{eqnarray*}
& & H^{(2)}=\sum_{j=1}^{N_1}\left[\sum_{i=1}^3\frac{p_{1ij}^2}{ 2m_1}+m_1gq_{1zj}\right]
 +\sum_{k=1}^{N_2}\sum_{i=1}^3\frac{p_{2ik}^2}{ 2m_2}\exp\left(\frac{2\alpha q_{2ik}}{ m_2}\right)\ \\ \
& & +\sum_{k=1}^{N_2}\left\{-\frac{gm_2^2}{ 2\alpha}\ln\left[1-\tanh^2\left(\sqrt\frac{\alpha}{ m_2g}~\frac{p_{2zk}}{
m_2}\right)
\right]+m_2gq_{2zk}\right\}\ .
\end{eqnarray*}
\begin{equation}
\label{III:H2T}\end{equation} 

Then, one can calculate for each Hamiltonian the canonical partition function (Toda et al 1998) which is associated to the
same statistical  system,

\begin{equation}
Z^{(i)}=\frac{1}{ N_1!N_2!h^{3N}}\int\exp\left(-\beta H^{(i)}\right)~dqdp\ \ i=1,2\ ,
\label{III:Z}\end{equation} 
where $\beta$ is defined as $\beta=1/kT$ with $k$ being the Boltzman's constant and T being the temperature, and the 
integration is carried out over all the coordinates and linear momenta of the two particles. The integration of momenta is
carried out in the intervale $(-\infty,+\infty)$. The integration on the transverse coordinates ($x,y$) is carried out in the
intervale $[0,L]$, and the integration of the vertical coordinate is carried out in the intervale $[0,z]$. The partition
functions for both cases are given by

\begin{eqnarray*}
Z^{(1)}=\frac{L^{2N_1}}{ N_1!N_2!h^{3N}}\left(\frac{2\pi m_1}{\beta}\right)^{3N_1/2} 
\left(\frac{1-e^{-\beta m_1gz}}{ \beta m_1g}\right)^{N_1}
\left(\frac{2\pi m_2}{\beta}\right)^{3N_2/2}\left(\frac{m_2}{\alpha}\right)^{2N_2}\times \ \\ \
(e^{\frac{-\alpha L}{m_2}}-1)^{2N_2}\Biggl[\sqrt{\frac{\pi}{2\beta\alpha g}}e^{-\frac{\beta m_2^2g}{2\alpha}}
\left(Erfi\left(\sqrt\frac{\beta gm_2^2}{ 2\alpha}~e^{-\alpha z/m_2}\right) -Erfi\left(\sqrt\frac{\beta g m_2^2}{
2\alpha}\right)\right)\Biggr]^{N_2}
\end{eqnarray*}
\begin{equation}
\label{III:Z1}\end{equation} 
and

\vfil\eject

\begin{eqnarray*}
Z^{(2)}&=&\frac{L^{2N_1}}{ N_1!N_2!h^{3N}}\left(\frac{2\pi m_1}{\beta}\right)^{3N_1/2} 
\left(\frac{1-e^{-\beta m_1gz}}{ \beta m_1g}\right)^{N_1}\left(\frac{2\pi m_2}{\beta}\right)^{N_2}
\left(\frac{m_2}{\alpha}\right)^{2N_2}\ \\ \ & & \times(e^{-\frac{\alpha L}{m_2}}-1)^{2N_2}
\left(\frac{1-e^{-\beta m_2gz}}{\beta m_2 g}\right)^{N_2}
\left[\sqrt\frac{\pi\alpha}{ m_2^3g}~\frac{\Gamma\left(\frac{\beta m_2^2g}{ 2\alpha}\right)}{
\Gamma\left(\frac{\beta m_2^2g}{ 2\alpha}+\frac{1}{ 2}\right)}\right]^{N_2}\ .
\end{eqnarray*}
\begin{equation}
\label{III:Z2}\end{equation} 
The system has two internal energies, $U^{(i)}=-\partial \ln Z^{(i)}/\partial\beta$,

\begin{equation}
U^{(1)}=\left(\frac{5N_1}{ 2}+{2N_2}\right)\frac{1}{\beta}-\frac{N_1m_1gze^{-\beta m_1gz}}{ 1-e^{-\beta m_1gz}}-\frac{N_2
m_2^2g}{ 2\alpha}-\frac{N_2 f'(\beta)}{ f(\beta)}
\label{III:U1}\end{equation} 
and

\begin{eqnarray*}
U^{(2)}&=&\left(\frac{5N_1}{ 2}+{2N_2}\right)\frac{1}{\beta}-
\frac{N_1m_1gze^{-\beta m_1gz}}{ 1-e^{-\beta m_1gz}}-\frac{N_2m_2gze^{-\beta m_2gz}}{ 1-e^{-\beta m_2gz}}\ \\ \
& & -\frac{N_2m_2^2g}{ 2\alpha}\left[\psi\left(\frac{\beta m_2^2g}{ 2\alpha}\right)-
\psi\left(\frac{\beta m_2^2g}{ 2\alpha}+\frac{1}{ 2}\right)\right]\ ,
\end{eqnarray*}
\begin{equation}
\label{III:U2}\end{equation} 
where the function $f(\beta)$ ($f'(\beta)=df(\beta)/d\beta$) has been defined as

\begin{eqnarray*}
f(\beta)&=&Erfi\left(\sqrt\frac{\beta gm_2^2}{ 2\alpha}~e^{-\alpha z/m_2}\right)
-Erfi\left(\sqrt\frac{\beta g m_2^2}{ 2\alpha}\right)\ ,
\end{eqnarray*}
\begin{equation}
\label{III:FB}\end{equation} 
and $Erfi(x)=-iErf(ix)$ is the complex error function which can be expressed in the form of the Dawson's integral, 
$Erfi(x)=\frac{2}{\sqrt{\pi}}e^{-x^2}Dawson(x)$,   $Dawson(x)=e^{-x^2}\int_0^{x}e^{t^2}$ and $\psi$ is the digamma function,
$\psi(x)=d\ln\Gamma(x)/dx$. Thus, one can have two heat capacity expressions for the system, $C_V^{(i)}=\partial
U^{(i)}/\partial T=-k\beta^2\partial U^{(i)}/\partial\beta$,
\begin{equation}
C_V^{(1)}=\left(\frac{5N_1}{ 2}+{2N_2}\right)k-\frac{N_1k\left(m_1gz\beta\right)^2
e^{-m_1gz\beta}}{ (1-e^{-m_1g z\beta})^2}+{N_2k\beta^2}\left(\frac{f''(\beta)}{ f(\beta)}-
\frac{(f'(\beta))^2}{ (f(\beta))^2}\right)
\label{III:CV1}\end{equation} 
and
\begin{eqnarray*}
& & C_V^{(2)}=\left(\frac{5N_1}{ 2}+{2N_2}\right)k
-\frac{N_1 k\left(m_1gz\beta\right)^2 e^{-m_1gz\beta}}{ (1-e^{-m_1g z\beta})^2}
-\frac{N_2 k\left(m_2gz\beta\right)^2 e^{-m_2gz\beta}}{ (1-e^{-m_2g z\beta})^2}\ \\ \
& &\hskip2cm +{N_2 k}\left(\frac{m_2^2g\beta}{ 2\alpha }\right)^2\left[\psi^{(1)}
\left(\frac{ m_2^2g\beta}{ 2\alpha }\right)-
\psi^{(1)}\left(\frac{m_2^2g\beta}{ 2\alpha }+\frac{1}{ 2}\right)\right]\ ,
\end{eqnarray*}
\begin{equation}
\label{III:CV2}\end{equation} 
where $\psi^{(1)}$ is the trigamma function, $\psi^{(1)}(x)=d^2\Gamma(x)/dx^2$. Figure 1 shows the difference 
$|C_V^{(1)}-C_V^{(2)}|$ as a function of $\beta=1/kT$. As one can see, this difference is not small at low temperatures (high
$\beta$ values). From $\beta$ lower than about 2100, $C_{V_2}$ is higher than $C_{V_1}$, and the situation is reversed for
higher values. This difference seems to have an important implication related with the ergodic hypothesis (Toda et al 1998).
Assuming the validity of the  hypothesis, one would expect not difference at all on the calculated heat capacities (or
internal energies) since averaging over the time variable must bring about the same value for both Hamiltonians (they
represent the same dynamical system). However, averaging over the canonical ensemble must be different if the Hamiltonians
are different. This ambiguity will remain when quantum canonical ensemble is considered (using Eq.~(\ref{II:E1}) and
Eq.~(\ref{II:E2})) for quantum statical analysis of the system. 

\section{\label{H} Conclusions }
We have shown two constants of motion, two Lagrangians, and two Hamiltonians for a free falling particle moving in a  media
with quadratic velocity dissipative force. These quantities describe the same dynamics of the classical system, but their
quantization and the associated statistical mechanics (for an ensemble of particles) describe two different quantum and
statistical dynamics. We have showed this at first order in the dissipative parameter and at first order in perturbation
theory. There is still a point which reamins to to study and has to deal with quasi-classical limit. The question is  whether
or not both quantum Hamiltonians, Eq.~(\ref{II:HQ}), describes the same quasi-classical dynamics ($\hbar\to 0$) and coincides
with the classical dynamics in this limit. We will deal with this problem  and hope to report some results soon.

{\bf Figure Captions}\\
Difference of the heat capacities as a function of $\beta=1/kT$ for $\alpha=0.01$, $g=1$, and $m_1/m_2=0.1$

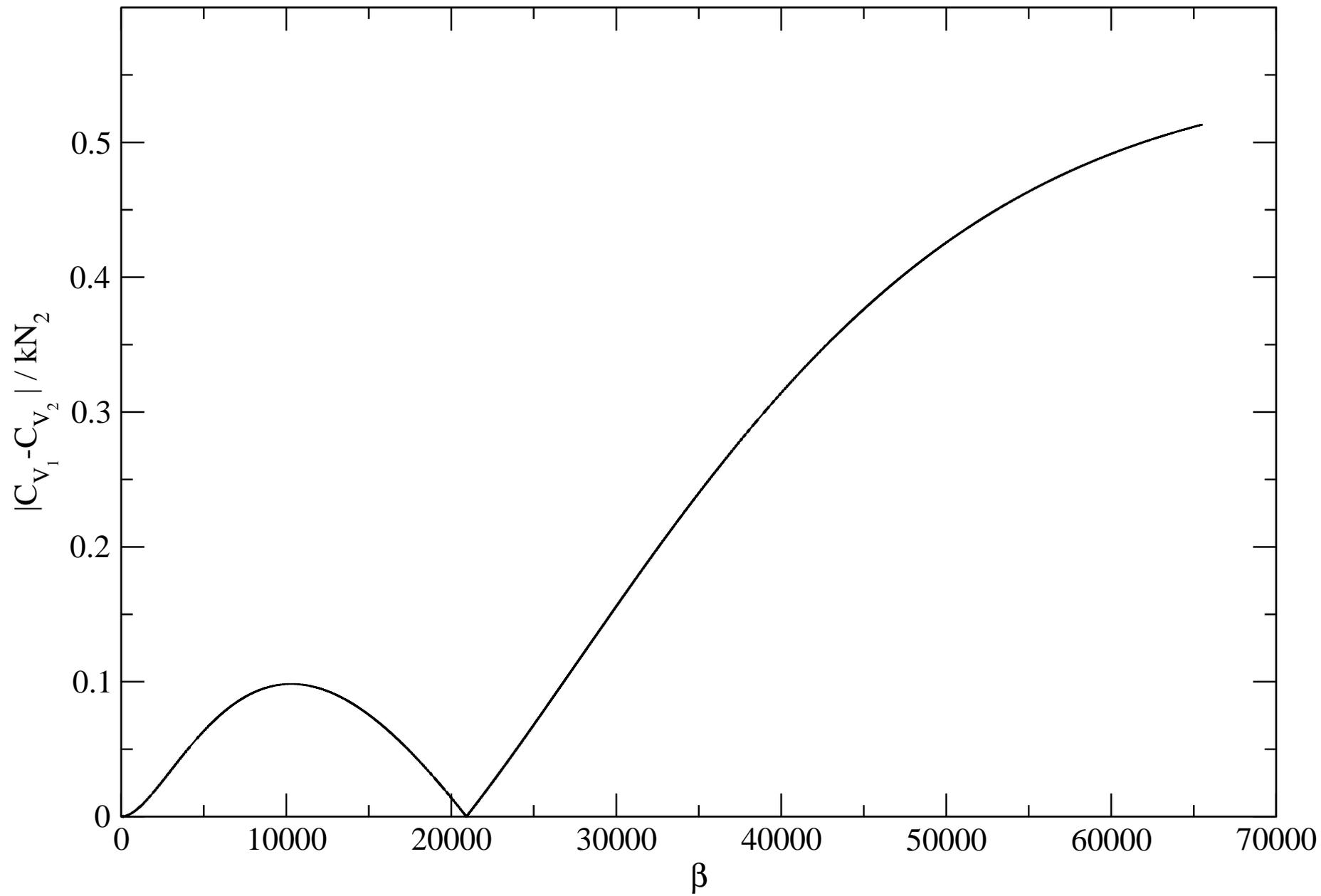
\end{document}